# SECURITY AND PRIVACY OF SENSITIVE DATA IN CLOUD COMPUTING: A SURVEY OF RECENT DEVELOPMENTS

Ali Gholami and Erwin Laure

HPCViz Dept., KTH- Royal Institute of Technology, Stockholm, Sweden
{gholami,erwinl}@pdc.kth.se

*ABSTRACT*

*Cloud computing is revolutionizing many ecosystems by providing organizations with computing resources featuring easy deployment, connectivity, configuration, automation and scalability. This paradigm shift raises a broad range of security and privacy issues that must be taken into consideration. Multi-tenancy, loss of control, and trust are key challenges in cloud computing environments. This paper reviews the existing technologies and a wide array of both earlier and state-of-the-art projects on cloud security and privacy. We categorize the existing research according to the cloud reference architecture orchestration, resource control, physical resource, and cloud service management layers, in addition to reviewing the existing developments in privacy-preserving sensitive data approaches in cloud computing such as privacy threat modeling and privacy enhancing protocols and solutions.*

## 1. INTRODUCTION

Cloud computing is revolutionizing many of our ecosystems, including healthcare. Compared with earlier methods of processing data, cloud computing environments provide significant benefits, such as the availability of automated tools to assemble, connect, configure and reconfigure virtualized resources on demand. These make it much easier to meet organizational goals as organizations can easily deploy cloud services. However, the shift in paradigm that accompanies the adoption of cloud computing is increasingly giving rise to security and privacy considerations relating to facets of cloud computing such as multi-tenancy, trust, loss of control and accountability [1]. Consequently cloud platforms that handle sensitive information are required to deploy technical measures and organizational safeguards to avoid data protection breakdowns that might result in enormous and costly damages.

Sensitive information in the context of cloud computing encompasses data from a wide range of different areas and disciplines. Data concerning health is a typical example of the type of sensitive information handled in cloud computing environments, and it is obvious that most individuals will want information related to their health to be secure. Hence, with the proliferation of these new cloud technologies in recent times, privacy and data protection requirements have been evolving to protect individuals against surveillance and database disclosure. Some examples of such protective legislation are the EU Data Protection Directive (DPD) [2] and the US Health Insurance Portability and Accountability Act (HIPAA) [3], both of which demand privacy preservation for handling personally identifiable information.

This paper presents an overview of the research on security and privacy of sensitive data in cloud computing environments. We identify new developments in the areas of orchestration, resource control, physical hardware, and cloud service management layers of a cloud provider. We also review the state-of-the-art in privacy-preserving sensitive data approaches for handling sensitive data in cloud computing such as privacy threat modeling and privacy enhancing protocols and solutions.

The rest of this paper is organized as follows. Section 2 gives an overview of cloud computing concepts and technologies. Section 3 describes the security and privacy issues that need to be solved in order to provide secure data management for cloud environments. Section 4, reviews the existing security solutions that are being used in the area of cloud computing. Section 5 describes research on privacy-preserving solutions for sensitive data. Finally, in Section 6, we present our findings and conclusions.

## 2. KEY CONCEPTS AND TECHNOLOGIES

Over the past few years, major IT vendors (such as Amazon, Microsoft and Google) have provided virtual machines (VMs), via their clouds, that customers could rent. These clouds utilize hardware resources and support live migration of VMs in addition to dynamic load-balancing and on-demand provisioning. This means that, by renting VMs via a cloud, the entire datacenter footprint of a modern enterprise can be reduced from thousands of physical servers to a few hundred (or even just dozens) of hosts.

While it is practical and cost effective to use cloud computing in this way, there can be issues with security when using systems that are not provided in-house. To look into these and find appropriate solutions, there are several key concepts and technologies that are widely used in cloud computing that need to be understood, such as virtualization mechanisms, varieties of cloud services, and "container" technologies.

### 2.1. Virtualization Mechanisms

A hypervisor or virtual machine monitor (VMM) is a key component that resides between VMs and hardware to control the virtualized resource [4]. It provides the means to run several isolated virtual machines on the same physical host. Hypervisors can be categorized into two groups [5]:

- *Type I*: Here the hypervisor runs directly on the real system hardware, and there is no operating system (OS) under it. This approach is efficient as it eliminates any intermediary layers. Another benefit with this type of hypervisor is that security levels can be improved by isolating the guest VMs. That way, if a VM is compromised, it can only affect itself and will not interfere with the hypervisor or other guest VMs.

- *Type II*: The second type of hypervisor runs on a hosted OS that provides virtualization services, such as input/output (IO) device support and memory management. All VM interactions, such as IO requests, network operations and interrupts, are handled by the hypervisor.

Xen[1] and kernel virtual machine (KVM)[2] are two popular open-source hypervisors (respectively of *Type I* and *Type II*). Xen runs directly on the underlying hardware and it inserts a virtualization layer between the system hardware and the virtual machines. The OSs running in the VMs interact with the virtual resources as if they were actually physical resources. KVM is a virtualization feature in the Linux Kernel that makes it possible to safely execute guest code directly on the host CPU.

**2.2. Cloud Computing Characteristics**

When considering cloud computing, we need to be aware of the types of services that are offered, the way those services are delivered to those using the services, and the different types of people and groups that are involved with cloud services.

Cloud computing delivers computing software, platforms and infrastructures as services based on pay-as-you go models. Cloud service models can be deployed for on-demand storage and computing power in various ways: *Software-as-a-Service* (*SaaS*), *Platform-as-a-Service (PaaS)* and *Infrastructure-as-a-Service (IaaS)*. Cloud computing service models have been evolved during the past few years within a variety of domains using the "*as-a-Service*" concept of cloud computing such as *Business Integration-as-a-Service, Cloud-Based Analytics-as-a-Service (CLAaaS), Data-as-a-Service (DaaS)* [61], [62]. This paper refers to the NIST cloud service models features [6] that are summarized in Table 1 that can be delivered to consumers using different models such as a private cloud, community cloud, public cloud, or hybrid cloud.

Table 1, Categorization of Cloud Service Models and Features

| Service Model | Function | Example |
|---|---|---|
| *SaaS* | Allows consumers to run applications by virtualizing hardware on the resources of the cloud providers | Salesforce Customer Relationship Management (CRM)[3] |
| *PaaS* | Provides capability of deploying custom applications with their dependencies within an environment called a container. | Google App Engine[4], Heroku[5] |
| *IaaS* | Provides a hardware platform as a service such as virtual machines, processing, storage, networks and database services. | Amazon Elastic Compute Cloud (EC2)[6] |

---

[1] Xen hypervisor, http://xen.org/products/xenhyp.html
[2] KVM, http://www.linux-kvm.org/
[3] Salesforce CRM, https://www.salesforce.com/crm/
[4] Google App Engine, https://appengine.google.com
[5] Heroku, https://www.heroku.com
[6] Amazon EC2, https://aws.amazon.com/ec2/

The NIST cloud computing reference architecture [7], defines five major actors in the cloud arena: cloud consumers, cloud providers, cloud carriers, cloud auditors and cloud brokers. Each of these actors is an entity (either a person or an organization) that participates in a cloud computing transaction or process, and/or performs cloud computing tasks.

A cloud consumer is a person or organization that uses services from cloud providers in the context of a business relationship. A cloud provider is an entity makes cloud services available to interested users. A cloud auditor conducts independent assessments of cloud services, operations, performance and security in relation to the cloud deployment. A cloud broker is an entity that manages the use, performance and delivery of cloud services, and also establishes relationships between cloud providers and cloud consumers. A cloud carrier is an entity that provides connectivity and transport of cloud services from cloud providers to cloud consumers through the physical networks.

The activities of cloud providers can be divided into five main categories: service deployment, resource abstraction, physical resources, service management, security and privacy [7]. Service deployment consists of delivering services to cloud consumers according to one of the service models (*SaaS, PaaS, IasS*). Resource abstraction refers to providing interfaces for interacting with networking, storage and compute resources. The physical resources layer includes the physical hardware and facilities that are accessible via the resource abstraction layer. Service management includes providing business support, resource provisioning, configuration management, portability and interoperability to other cloud providers or brokers. The security and privacy responsibilities of cloud providers include integrating solutions to ensure legitimate delivery of cloud services to the cloud consumers. The security and privacy features that are necessary for the activities of cloud providers are described in Table 2 [10].

Table 2, Security and Privacy Factors of the Cloud Providers

| Security Context | Description |
| --- | --- |
| **Authentication and Authorization** | Authentication and authorization of cloud consumers using pre-defined identification schemes. |
| **Identity and Access Management** | Cloud consumer provisioning and deprovisioning via heterogeneous cloud service providers. |
| **Confidentiality, Integrity, Availability** | Assuring the confidentiality of the data objects, authorizing data modifications and ensuring that resources are available when needed. |
| **Monitoring and Incident Response** | Continuous monitoring of the cloud infrastructure to assure compliance with consumer security policies and auditing requirements. |
| **Policy Management** | Defining and enforcing rules for certain actions such as auditing or proof of compliance. |
| **Privacy** | Protect personally identifiable information (PII) within the cloud from adversarial attacks that aim to find out the identity of the person that PII relates to. |

The majority of cloud computing infrastructures consist of reliable services delivered through data centers to achieve high availability through redundancy. A data center or computer center is a facility used to house computer systems and associated components, such as storage and network systems. It generally includes redundant or backup power units, redundant network connections, air conditioning, and fire safety controls.

**2.3. Containers Technology**

Clouds based on Linux container (LXC) technology are considered to be next-generation clouds, so LXCs has become an important part of the cloud computing infrastructures because of their ability to run several OS-level isolated VMs within a host with a very low overhead. LXCs are built on modern kernel features. An LXC resembles a light-weight execution environment within a host system that runs instructions native to the core CPU while eliminating the need for instruction level emulation or just-in-time compilation [8]. LXCs contain applications, configurations and the required storage dependencies, in a manner similar to the just enough OS (JeOS).

Using containers, several applications can share an OS, binaries or libraries, which results in significant increases in efficiency compared to using hypervisors. For example, the portability of applications and the provisioning time of VMs are very low with container technologies [9].

LXC technologies were introduced in the 1980s, starting with the chroot (change root) command, and evolving into to popular container managers such as Docker.

- **Chroot**: The Unix chroot system call, which was introduced as part of Unix version 7 in 1979, can be considered as the first step in the evolution of containerization. The chroot call changes the root directory of the calling process to a specified path, where the root directory is known by all children of the calling process. This feature is used by some containers for isolation and sharing the underlying file system. Chroot is often used when building system images by changing root to a temporary directory, downloading and installing packages in chroot or compressing chroot as a system root file system.

- **FreeBSD Jail**[7] extended chroot in 1998 to provide enhanced security. FreeBSD jail settings can explicitly restrict access outside the sandbox environment by files, processes, and user accounts (including accounts created by the jail definition). Jail can therefore define a new root user, who has full control inside the sandbox, but who cannot reach anything outside.

- **Namespaces** were introduced in 1992 [11] for process-based resource isolation. Namespaces provide tools for isolating the view of global resources such as details about file systems, processes, network interfaces, Inter Process Communication (IPC), host names, and user IDs. Processes in a particular namespace are invisible to other processes because they think that they are the only processes on the system and because "connectivity" is only permitted with the parent namespace

---

[7] https://www.freebsd.org/doc/handbook/jails.html

- **Control Groups (cgroups)[8]** are kernel mechanisms introduced by Google in 2007 to provide fine-grained control by grouping processes and their children into a tree structure for resource management. Each group can be assigned a task for operations related to CPU, memory, disk and network. For example, to isolate two groups such as applications resources and OS resources, two groups (group 1 and 2) can be created to assign resource profiles to each group.

- **Linux Security Modules (LSMs)** are kernel modules which provide a framework for mandatory access control (MAC) security implementations. In MAC implementations, the administrator (user or process) assigns access controls to subject / initiator. In discretionary access control (DAC), the resource owner (user) assigns access controls to the subject or initiator. Existing LSM implementations include AppArmor, SELinux and so forth to prevent virtual machines from attacking other virtual machines or the host. For this purpose, policies are used to define what actions a process can perform on a particular system.

- **Containers** are built on the hardware and operating system but they make use of kernel features called chroots, cgroups and namespaces to construct a contained environment without the need for a hypervisor. The most recent container technologies are Solaris Zones, OpenVZ and LXC.

  In 2004, Solaris version 10 used zones as facilities to provide protected virtualized environments within a single host. Every Solaris system includes a global zone for both system and system-wide administrative control, and may have one or more non-global zones. All processes run in the global zone if there is no non-global zone. The global zone is aware of all devices and all file systems, while non-global zones are not aware of the existence of any other zones. Zone-based containers provide isolation, security and virtualization. Zones are similar to jails with additional features such as snapshots and cloning that make it possible to clone efficiently or to duplicate a current zone into a new zone.

  In 2005 OpenVZ [9] containers were introduced using a modified Linux kernel with a set of extensions. OpenVZ is based on the namespace and control group concepts in contrast to jails, which were used in FreeBSD.

  Later in 2008, LXC[10] emerged as a container management tool and it combined namespaces and control groups to create a fully isolated environment. It provides libraries and command-line support to enable administrators to create new containers. LXC containers can be used in either privileged (as a root user) or unprivileged (as a non-root user) modes to easily customize kernel capabilities or configure cgroups to satisfy the particular requirements.

  Docker is another container management tool – it was introduced in 2013 and is based on namespaces, cgroups and SELinux. Docker provides automation for the deployment of containers through remote APIs and has additional features that make it possible to create standardized environments for developing applications. This has made Docker a popular technology. Creating the standardized environments is achieved using a layered image format that enables users to add or remove applications and their dependencies to form a trusted image.

---

[8] Kernel CGroups, https://www.kernel.org/doc/Documentation/cgroups/
[9] OpenVZ, https://openvz.org/
[10] LXC Containers, https://linuxcontainers.org/

Docker adds portable deployment of LXCs across different machines. In cloud terms, one can think of LXC as the hypervisor and Docker as both the open virtualization appliance and the provision engine [12]. Docker images can run unchanged on any platform that supports Docker. In Docker, containers can be created from build files such as Web service management.

The use of containers in cloud computing is increasingly becoming popular amongst cloud providers such as Google[11] and Microsoft[12]. Significant improvements in performance and security are the main driving factors for employing containers compared to virtualization using hypervisors in cloud infrastructures.

## 3. CLOUD SECURITY AND PRIVACY CHALLENGES

Cloud computing has raised several security threats such as data breaches, data loss, denial of service, and malicious insiders that have been extensively studied in [67], [68]. These threats mainly originate from issues such as multi-tenancy, loss of control over data and trust. (Explanations of these issues follow in the next subsection.)

Consequently the majority of cloud providers – including Amazon's Simple Storage Service (S3)[13], the Google Compute Engine[14] and the Citrix Cloud Platform[15] - do not guarantee specific levels of security and privacy in their service level agreements (SLAs) as part of the contractual terms and conditions between cloud providers and consumers. This means that there are important concerns related to security and privacy that must be taken into consideration in using cloud computing by all parties involved in the cloud computing arena. These are discussed in the subsection 2.2.

### 3.1. Security Issues in Cloud Computing

- **Multi-tenancy:** Multi-tenancy refers to sharing physical devices and virtualized resources between multiple independent users. Using this kind of arrangement means that an attacker could be on the same physical machine as the target. Cloud providers use multi-tenancy features to build infrastructures that can efficiently scale to meet customers' needs, however the sharing of resources means that it can be easier for an attacker to gain access to the target's data.

- **Loss of Control:** Loss of control is another potential breach of security that can occur where consumers' data, applications, and resources are hosted at the cloud provider's owned premises. As the users do not have explicit control over their data, this makes it possible for cloud providers to perform data mining over the users' data, which can lead to security issues. In addition, when the cloud providers backup data at different data centers, the consumers cannot be sure that their data is completely erased everywhere when they delete their data. This has the potential to lead to misuse of the unerased data. In these types of situations where the consumers lose control over their data, they see the cloud provider as a black-box where they cannot directly monitor the resources transparently.

---
[11] Google Container Engine, https://cloud.google.com/container-engine/
[12] Microsoft Azure Container, https://azure.microsoft.com/en-us/blog/azure-container-service-now-and-the-future/
[13] Amazon S3 SLA, https://aws.amazon.com/s3/sla/
[14] Google Compute Engine SLA, https://cloud.google.com/compute/sla
[15] Citrix Cloud Platform SLA, https://www.citrix.se/products/cloudplatform/overview.html

- **Trust Chain in Clouds:** Trust plays an important role in attracting more consumers by assuring on cloud providers. Due to loss of control (as discussed earlier), cloud users rely on the cloud providers using trust mechanisms as an alternative to giving users transparent control over their data and cloud resources. Therefore cloud providers build confidence amongst their customers by assuring them that the provider's operations are certified in compliance with organizational safeguards and standards.

### 3.2. Privacy Considerations of Processing Sensitive Data

The security issues in cloud computing lead to a number of privacy concerns. Privacy is a complex topic that has different interpretations depending on contexts, cultures and communities, and it has been recognized as a fundamental human right by the United Nations [13]. It worth nothing that privacy and security are two distinct topics although security is generally necessary for providing privacy [1], [59].

Several efforts have been made to conceptualize privacy by jurists, philosophers, researchers, psychologists, and sociologists in order to give us a better understanding of privacy – for example, Alan Westin's research in 1960 is considered to be the first significant work on the problem of consumer data privacy and data protection. Westin [14] defined privacy as follow.

"*Privacy is the claim of individuals, groups, or institutions to determine for themselves when, how, and to what extent information about them is communicated to others.*"

The International Association of Privacy Professionals (IAPP)[16] glossary 27 refers to privacy as the appropriate use of information under the circumstances. The notion of what constitutes appropriate handling of data handling varies depending on several factors such as individual preferences, the context of the situation, law, collection, how the data would be used and what information would be disclosed.

In jurisdictions such as the US, "privacy" is the term that is used to encompass the relevant laws, policies and regulations, while in the EU the term "data protection" is more commonly used when referring to privacy laws and regulations. Legislation that aims to protect the privacy of individuals – such as the European Union (EU) DPD [1], the Gramm-Leach-Bliley Act (GLBA) [15], the Right to Financial Privacy Act (RFPA) [16], and the HIPAA [2] – can become very complicated and have a variety of specific requirements. Organizations collecting and storing data in clouds that are subject to data protection regulations must ensure that the privacy of the data is preserved appropriately to lay the foundations for legal access to sensitive personal data.
The development of a legal definition for cybercrime, the issue of jurisdiction (who is responsible for what information and where are they held responsible for it) and the regulation of data transfers to third countries [17] are among other challenging issues when it comes to security in cloud computing. For example, the DPD, which is the EU's initial attempt at privacy protection, presents 72 recitals and 34 articles to harmonize the regulations for information flow within the EU Member States.

---

[16] IAPP Glossary, https://iapp.org/resources/glossary

The DPD highlights the demand for cross-border transfer of data through non-legislative measures and self-control. One example of where these types of privacy principles are being used is the Safe Harbor Agreement (SHA) which makes it possible transfer data to US-based cloud providers that are assumed to have appropriate data protection mechanisms. However, cloud carriers are not subject to the SHA, which leads to complexity in respect to international laws.

There is an ongoing effort [18] to replace the EU DPD with a new a data protection regulation containing 91 articles that aims to lay out a data protection framework in Europe. The proposed regulations expand the definition of personal data protection to cover any information related to the people who are the subjects of the data, irrespective of whether the information is private, public or professional in nature. The regulations also include definitions of new roles related to handling data (such as data transfer officers) and propose restricting the transfer of data to third-party countries that do not guarantee adequate levels of protection. Currently Argentina, Canada, Guernsey, Jersey, the Isle of Man, Israel, Switzerland, the US Safe Harbor Privacy Program, and the US Transfer of Air Passenger Name Data Record are considered to offer adequate protection. The new regulations consider imposing significant penalties for privacy breaches that result from violations of the regulations, for example, such a penalty could be 0.5 percent of the worldwide annual turnover of the offending enterprise.

## 4. SECURITY SOLUTIONS

This section reviews the research on security solution such as authentication, authorization, and identity management that were identified in Table 2.2 [10] as being necessary so that the activities of cloud providers are sufficiently secure.

### 4.1 Authentication and Authorization

In [19] the authors propose a credential classification and a framework for analyzing and developing solutions for credential management that include strategies to evaluate the complexity of cloud ecosystems. This study identifies a set of categories relevant for authentication and authorization for the cloud focusing on infrastructural organization which include classifications for credentials, and adapt those categories to the cloud context. The study also summarizes important factors that need to be taken into consideration when adopting or developing a solution for authentication and authorization – for example, identifying the appropriate requirements, categories, services, deployment models, lifecycle, and entities. In other work, a design model for multi-factor authentication in cloud computing environments is proposed in [20], and this model includes an analysis of the potential security threats in the proposed model. Another authentication solution is seen with MiLAMob [21], which provides a SaaS authentication middleware for mobile consumers of IaaS cloud applications. MiLAMob is a middleware-layer that handles the real-time authentication events on behalf of consumer devices with minimal HTTP traffic. The middleware currently supports mobile consumption of data on IaaS clouds such as Amazon's S3.

FermiCloud [22] uses another approach for authentication and authorization - it utilizes public key infrastructure (PKI) X.509 certificates for user identification and authentication. FermiCloud is built in OpenNebula1[17] and it develops both X.509 authentication in Sunstone OpenNebula – a Web interface intended for user management – and X.509 authentication via command-line interfaces. To avoid the limitations of OpenNebula access control lists that are used for authorization after successful authentication of users, authors integrated an existing local credential mapping service. This solution has also been extended in cloud federations to authorize users across different cloud providers that have established trust relationships through trusted certification authorities.

Tang et al. [23] introduce collaborative access control properties such as centralized facilities, agility, homogeneity, and outsourcing trust. They have introduced an authorization-as-a-service (AaaS) approach using a formalized multi-tenancy authorization system, and providing administrative control over enhanced fine-grained trust models. Integrating trust with cryptographic role-based access control (RBAC) [24] is another solution that ensures trust for secure sharing of data in the cloud. The authors propose using cryptographic RBAC to enforce authorization policies regarding the trustworthiness of roles that are evaluated by the data owner. Another feature of the authorization system in this solution is that it develops a new concept using role inheritance for evaluating the trustworthiness of the system. In another study, Sendo et al. [25] propose a user-centric approach for platform-level authorization of cloud services using the OAuth2 protocol to allow services to act on behalf of users when interacting with other services in order to avoid sharing usernames and passwords across service.

### 4.2 Identity and Access Management

The important functionalities of identity management systems for the success of clouds in relation to consumer satisfaction is discussed in [26]. The authors also present an authorization system for cloud federation using Shibboleth - an open source implementation of the security assertion markup language (SAML) for single sign-on with different cloud providers. This solution demonstrates how organizations can outsource authentication and authorization to third-party clouds using an identity management system. Stihler et al. [27] also propose an integral federated identity management for cloud computing. A trust relationship between a given user and SaaS domains is required so that SaaS users can access the application and resources that are provided. In a PaaS domain, there is an interceptor that acts as a proxy to accept the user's requests and execute them. The interceptor interacts with the secure token service (STS), and requests the security token using the WS-Trust specification.

IBHMCC [28] is another solution that contains identity-based encryption (IBE) and identity-based signature (IBS) schemes. Based on the IBE and IBS schemes, an identity-based authentication for cloud computing has been proposed. The idea is based on the identity-based hierarchical model for cloud computing along with the corresponding encryption and signature schemes without using certificates for simplified key management.

Contrail [29] is another approach that aims to enhance integration among heterogeneous clouds both vertically and horizontally. Vertical integration provides a unified platform for the different kinds of resources while horizontal integration abstracts the interaction models of different cloud providers. In [29] the authors develop a horizontal federation scheme as a requirement for vertical integration. The proposed federation architecture contains several layers, such as users' identities, business logic and a federation manager to support APIs for resources, storage, and networking across different providers.

---

[17] http://opennebula.org/

E-ID authentication and uniform access to cloud storage service providers [30] is an effort to build identity management systems for authenticating Portuguese citizens using national e-identification cards for cloud storage systems. In this approach, the OAuth protocol is integrated for authorizing the cloud users. The e-ID cards contain PKI certificates that are signed by several levels of governmental departments. A certification authority is responsible for issuing the e-ID cards and verifying them. The e-ID cards enable users for identity-based encryption of data in cloud storage.

In [31], the authors consider the issues related to inter-cloud federation and the proposed ICEMAN identity management architecture. ICEMAN discusses identity life cycle, self-service, key management, provisioning and deprovisioning functionalities that need to be included in an appropriate intercloud identity management system.

The EGI delivered a hybrid federated cloud [32] as a collaboration of communities developing, innovating, operating and using clouds for research and education. The EGI federated cloud provides IaaS, persistent block storage attached to VMs, and object-level storage for transparent data sharing. The EGI controls access to resources using X.509 certificates and the concept of "Virtual Organization" (VO). VO refers to a dynamic set of users or institutions using resource-sharing rules and conditions. The authorization attributes are issued through a VO management system that can be integrated with SAML for federation.

### 4.3 Confidentiality, Integrity, and Availability

Santos et al. [33] extend the Terra [34] design that enables users to verify the integrity of VMs in the cloud. The proposed solution is called the trusted cloud computing platform (TCCP), and the whole IaaS is considered to be a single system instead of granular hosts in Terra. In this approach, all nodes run a trusted virtual machine monitor to isolate and protect virtual machines. Users are given access to cloud services through the cloud manager component. The external trusted entity (ETE) is another component that provides a trust coordinator service in order to keep track of the trusted VMs in a cluster. The ETE can be used to attest the security of the VMs. A TCCP guarantees confidentiality and integrity in data and computation and it also enables users to attest to the cloud service provider to ensure whether the services are secure prior to setting up their VMs. These features are based on the trusted platform module (TPM) chip. The TPM contains a private endorsement key that uniquely identifies the TPM and some cryptographic functions that cannot be altered.

In 2011, Popa et al. proposed CloudProof [35] as a secure storage system to guarantee confidentiality, integrity and write-serializability using verifiable proofs of violation by external third parties. Confidentiality is ensured by private keys that are known only to the owner of the data that is to be encrypted. The main idea behind CloudProof is the use of the attestation mechanism. Attestations provide proof of sanity of users, data owners and cloud service providers. Data owners use a block identifier to acquire the content of a block. This mechanism enables users to store data by putting a block identifier and the contents of the block in the cloud. The attestation structure implements a solution called "block hash" for performing integrity checks through signature verification. The block hash provides proof for write-serializabilty using a forked sequence of the attestations while a chain hash is used for a broken chain of attestations which are not sequenced correctly.

Fuzzy authorization (FA) for cloud storage [36] is another flexible and scalable approach to enable data to be shared securely among cloud participants. FA ensures confidentiality, integrity and secure access control by utilizing secret sharing schemes for users with smartphones who are using the cloud services.

In [37] the authors define threats to cloud server hypervisors thorough analysis of the codebase of two popular open-source hypervisors: Xen and KVM. In addition, they discuss the vulnerabilities reports associated with them. As a result, a model is proposed for characterization of hypervisor vulnerabilities in three dimensions: the trigger source, the attack vector and the attack target. The attack vector consists of the Hypervisor functionality that makes security breaches possible - for example, virtual CPUs, symmetric multiprocessing, soft memory management units, interrupt and timer mechanisms, IO and networking, paravirtualized IO, VM exits, hypercalls, VM management (configure, start, pause and stop VMs), remote management, and software hypervisor add-ons. Successful exploitation of a vulnerability in these functionalities enables an attacker to compromise the confidentiality, integrity, or availability of the Hypervisor or one of its guest VMs.

The vulnerability reports in [37] show 59 vulnerability cases for Xen and 38 cases for KVM. Approximately 50 percent of these vulnerabilities are the same for Xen a dKVM and consist of issues relating to confidentiality, integrity and availability. The remote management software of Xen contributes to 15.3 percent of the vulnerabilities that demonstrates the increase attack surface by non-essential services. The VM management component contributes to 11.9 percent of the vulnerabilities in Xen compared to 5.3 percent in KVM. The lower vulnerability rate in KVM is due to the libvirt toolkit inside the hypervisor, whereas Xen's decision to allocate an entire privileged is done in Dom0. Other factors that have been studied in [20] are trigger sources and likely attack targets, including the overall network, the guest VM's user-space, the guest VM's kernel-space, the Dom0/host OS, and the hypervisor. The most common trigger source is the guest VM user-space, which gives rise to 39.0 percent of Xen's and 34.2 percent of KVM's vulnerabilities. This makes it possible for any user-space guest VM to be a threat to the hypervisor. The guest VM kernel-space has around 32 percent of the total vulnerabilities in both cases. The authors show Dom0 to be a more common target than the hypervisor in Xen, whereas the host OS in KVM is a less common target compared to the hypervisor. The location of the IO device emulation back-end drivers plays an important factor in this difference. The IO and network device emulation functionalities cause one third of the 15 vulnerabilities in both.

In [38] the authors propose Swap and Play as a new approach for live updating of hypervisors without the need to reboot the VM for high availability. The proposed design is scalable, usable and applicable in cloud environments and it has been implemented in Xen as one of the most popular hypervisors. Swap and Play provides methods to transfer the in-memory state of the running hypervisor to the updating state, in addition to updating the underlying host. Swap and Play consists of three independent phases: preparation, distribution and update. In the preparation phase information for the later state transfer is collected. The distribution phase deploys the update package on the target host for updating. In the last step, the update package is patched to individual hosts in the cloud. Each host applies the update package independently of the others and does not require any network resources. The Xen implementation of the Swap and Play solution is called SwapVisor. SwapVisor introduces a new hypercall in the Xen architecture. A hypercall is a trap from a domain to the hypervisor (similar to a syscall from an application to the kernel). Hypercalls are used by domains to request privileged operations such as updating page tables. The experiments show that updating from Xen version 4.2.0 to version 4.2.1 is fulfilled within approximately 45 ms which seems to be intangible and have almost zero effect on the network performance.

Klein et al. [39] improve cloud service resilience using a load-balancing mechanism called brownout. The idea behind this solution is to maximize the optional contents to provide a solution that is resilient to volatility in terms of flash crowds and capacity shortages (through load-balancing over replicas) when compared to other approaches that are implemented using response-time or queue length. In another effort [40] the authors proposed a synchronization mechanism for cloud accounting systems that are distributed. The run time resource usage generated from different clusters is synchronized to maintain a single cloud-wide view of the data so that a single bill can be created. The authors also proposed a set of accounting system requirements and an evaluation method which verifies that the solution fulfills these requirements.

## 4.4 Security Monitoring and Incident Response

Anand [41] presents a centralized monitoring solution for cloud applications consisting of monitoring the server, monitors, agents, configuration files and notification components. Redundancy, automatic healing, and multi-level notifications are other benefits of the proposed solution which are designed to avoid the typical drawbacks of a centralized monitoring system, such as limited scalability, low performance and single point of failure.

Brinkmann et al. [42] present a scalable distributed monitoring system for clouds using a distributed management tree that covers all the protocol-specific parameters for data collection. Data acquisition is done through specific handler implementations for each infrastructure-level data supplier. Data suppliers provide interoperability with cloud software, virtualization libraries and OS-level monitoring tools. The authors review the limitations of existing intrusion detection systems and discuss VM-level intrusion detection as an emerging area for securing VMs in cloud environments. The requirements for an efficient intrusion detection system for cloud infrastructures – including multi-tenancy, scalability and availability – are identified and a VM introspection detection mechanism via a hypervisor is proposed.

Hypervisor-based cloud intrusion detection systems are a new approach (compared to existing host-based and network-based intrusion detection systems) that is discussed in [43]. The idea is to use hypervisor capabilities to improve performance over data residing in a VM. Performance metrics are defined as networking transmitted and received data, read/write over data blocks, and CPU utilization. These metrics are retrieved in near real-time intervals by endpoint agents that are connected directly to a controller that analyzes the collected data using signatures to find any malicious activity. The controller component sends an alert to a notification service in case there is any potential attack.

## 4.5 Security Policy Management

In [44] the authors propose a generic security management framework allowing providers of cloud data management systems to define and enforce complex security policies through a policy management module. The user activities are stored and monitored for each storage system, and are made available to the policy management module. Users' actions are evaluated by a trust management module based on their past activities and are grouped as "fair" or "malicious". An appropriate architecture for security management which satisfies the requirements of policy definitions (such as flexibility, expressiveness, extendibility and correctness) has been implemented. The authors evaluated the proposed system on a data management system that is built on data storage.

Takabi et al. [45] introduce policy management as a service (PMaaS) to provide users with a unified control point for managing access policies in order to control access to cloud resources independently of the physical location of cloud providers. PMaaS is designed specifically to solve the issue of having multiple access control authorization mechanisms employed by cloud service providers that restrict the flexibility of applying custom access control to a particular service. For this purpose, the PMaaS architecture includes a policy management service provider that is the entry point for cloud users to define and manage the policies. The cloud service provider imports the user-defined policies and acts a policy decision point to enforce the user policies.

The challenges associated with policy enforcement in heterogeneous distributed environments are discussed in [46]. The authors propose a framework to support flexible policy enforcement and a feedback system using rule- and context-based access control to inform cloud users about the effect of defined policies. There are three main requirements for building a general policy enforcement framework. First it must support various data types such as image, structured and textual data. Secondly, in a distributed environment there need to be several compute engines such as Map/Reduce, relational database management systems or clusters. Finally, access policy requirements in terms of access control policies, data sharing policies, and privacy policies need to be integrated with the general policy management framework. Several policy enforcement mechanisms (such as extensible access control markup language or inline-reference monitors to enforce user-centric policies in accord with cloud provider approval) were also discussed.

In [47] the authors describe A4Cloud with the aim of developing solutions to ensure accountability and transparency in cloud environments. Users need to be able to track their data usage to know how the cloud provider satisfies their expectations for data protection. For this purpose cloud providers must employ solutions that provide users with appropriate control and transparency over their data, e.g. tools to define policies for compliance with regulatory frameworks. In another effort [48] the authors discuss the issue of usable transparent data processing in cloud computing and also consider how to enable users to define transparency policies over their data. They identify the requirements for transparent policy management in the cloud based on two aspects: user demands and legal aspects of transparent data processing.

# 5. PRIVACY-PRESERVATION FOR SENSITIVE DATA IN CLOUD COMPUTING

Over the time, organizations have collected valuable information about the individuals in our societies that contain sensitive information, e.g. medical data. Researchers need to access and analyze such data using big data technologies [63], [64], [65] in cloud computing, while organizations are required to enforce data protection compliance (subsection 3.2).

There has been considerable progress on privacy preservation for sensitive data in both industry and academia, e.g., solutions that develop protocols and tools for anonymization or encryption of data for confidentiality purposes. This section categorizes work related to this area according to different privacy protection requirements. However, these solutions have not yet been widely adopted by cloud service providers or organizations.

Pearson [1] discusses a range of security and privacy challenges that are raised by cloud computing. Lack of user control, lack of training and expertise, unauthorized secondary usage, complexity of regulatory compliance, transborder data flow restrictions and litigation are among the challenges faced in cloud computing environments. In [66], the authors describe the privacy challenges of genomic data in the cloud including terms of services of cloud providers that are not developed with a healthcare mindset, awareness of patient to upload their data into the cloud without their consent, multi-tenancy, data monitoring, data security and accountability. The authors also provide recommendations for data owners when aiming to use cloud provider services.

In [49] the authors discussed several privacy issues associated with genomic sequencing. This study also described several open research problems (such as outsourcing to cloud providers, genomic data encryption, replication, integrity, and removal of genomic data) along with giving suggestions to improve privacy through collaboration between different entities and organizations. In another effort [50], raw genomic data storage through encrypted short reads is proposed.

Outsourcing privacy is another topic that is discussed in [51]. The authors define the concept of "outsourcing privacy" where a database owner updates the database over time on untrusted servers. This definition assumes that database clients and the untrusted servers are not able to learn anything about the contents of the databases without authorized access. The authors implements a server-side indexing structure to produce a system that allows a single database owner to privately and efficiently write data to, and multiple database clients to privately read data from, an outsourced database.

Homomorphic encryption is another privacy-preserving solution that is based on the idea of computing over encrypted data without knowing the keys belonging to different parties. To ensure confidentiality, the data owner may encrypt data with a public key and store data in the cloud. When the process engine reads the data, there is no need to have the DP's private key to decrypt the data. In private computation on encrypted genomic data [52], the authors proposed a privacy-preserving model for genomic data processing using homomorphic encryption on genome-wide association studies.

Anonymization is another approach to ensure the privacy of sensitive data. SAIL [53] provides individual-level information on the availability of data types within a collection. Researchers are not able to cross-link (which is similar to an equality join in SQL) data from different outside studies, as the identities of the samples are anonymized. In another effort [57] the authors propose an integration architecture to make it possible to perform aggregated queries over anonymized medical data sets from different data providers. In this solution, data providers remove the data subjects' identifiers and apply a two-level encryption using hashing and PKI certificates. The sensitive information will then be anonymized using an open-source toolkit and will be encrypted granularly using the cloud provider's public key. ScaBIA [60] is another solution for processing and storing anonymized brain imaging data in cloud. This approach provides PKI authentication for administrator roles to deploy a PaaS middleware and defines researchers as users in the in Microsoft Azure cloud. Researchers are allowed to login by username/password to run statistical parametric mapping workflows within isolated generic worker containers. The brain imaging datasets and related results can be shared by the researchers using a RBAC model over secure HTTPS connections.

In [54], the design and implementation of a security framework for BiobankCloud, a platform that supports the secure storage and processing of genomic data in cloud computing environments, has been discussed. The proposed framework is built on the cloud privacy threat modeling approach [55], [56] which is used to define the privacy threat model for processing next-generation sequencing data according to the DPD [2]. This solution includes a flexible two-factor authentication and an RBAC access control mechanism, in addition to auditing mechanisms to ensure that the requirements of the DPD are fulfilled.

## 6. CONCLUSIONS

This paper surveyed recent advances in cloud computing security and privacy research. It described several cloud computing key concepts and technologies, such as virtualization, and containers. We also discussed several security challenges that are raised by existing or forthcoming privacy legislation, such as the EU DPD and the HIPAA.

The results that are presented in the area of cloud security and privacy are based on cloud provider activities, such as providing orchestration, resource abstraction, physical resource and cloud service management layers. Security and privacy factors that affect the activities of cloud providers in relation to the legal processsioning of consumer data were identified and a review of existing research was conducted to summarize the state-of-the-art in the field.

### ACKNOWLEDGEMENTS

This work is funded by the EU FP7 project "Scalable, Secure Storage and Analysis of Biobank Data" under Grant Agreement no. 317871.

## AUTHORS

**Ali Gholami** is a PhD student at the KTH Royal Institute of Technology. His research interests include the use of data structures and algorithms to build adaptive data management systems. Another area of his research focuses on the security concerns associated with cloud computing. He is currently exploring strong and usable security factors to enable researchers to process sensitive data in the cloud.

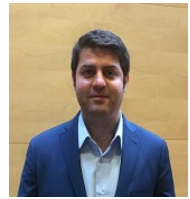

**Professor Erwin Laure** is Director of the PDC - Center for High Performance Computing Center at KTH, Stockholm. He is the Coordinator of the EC-funded "EPiGRAM" and "ExaFLOW" projects as well as of the HPC Centre of Excellence for Bio-molecular Research "BioExcel" and actively involved in major e-infrastructure projects (EGI, PRACE, EUDAT) as well as exascale computing projects. His research interests include programming environments, languages, compilers and runtime systems for parallel and distributed computing, with a focus on exascale computing.

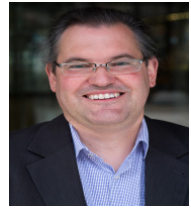